\documentclass{elsart}
\usepackage{epsfig,amssymb}
\begin{document}

\begin{frontmatter}

\title{Testing Option Pricing with the Edgeworth Expansion}

\author[itau]{ Ruy Gabriel Balieiro Filho} \ead{ruy.balieiro-filho@itau.com.br} 
and
\author[ift]{Rogerio Rosenfeld } \footnote{Talk given by R. Rosenfeld at the
{\sl International Workshop on Trends and Perspectives on Extensive and Non-Extensive Statistical Mechanics},
19--21 November, Angra dos Reis, Brazil } \ead{rosenfel@ift.unesp.br}

\address[itau]{Banco Itau \\
Av. Engenheiro Arruda Pereira, 707 - 04344-902
S\~{a}o Paulo, SP, Brazil}

\address[ift]{Instituto de F\'{\i}sica Te\'orica - UNESP\\
Rua Pamplona, 145 - 01405-900, S\~{a}o Paulo, SP, Brazil}

\vspace{0.1cm}

\begin{abstract}
There is a well developed framework, the Black-Scholes theory, for the pricing of contracts based on the future
prices of certain assets, called options. This theory assumes that the probability distribution of the returns
of the underlying asset is a gaussian distribution. However, it is observed in the market that this hypothesis
is flawed, leading to the introduction of a fudge factor, the so-called volatility smile. Therefore, it would be
interesting to explore extensions of the Black-Scholes theory to non-gaussian distributions.  In this
contribution we provide an explicit formula for the price of an option when the distributions of the returns of
the underlying asset is parametrized by an Edgeworth expansion, which allows for the introduction of higher
independent moments of the probability distribution, namely skewness and kurtosis. We test our formula with options
in the brazilian and american markets, showing that the volatility smile can be reduced. We also check whether our
approach leads to more efficient hedging strategies of these instruments.
\end{abstract}

\end{frontmatter}

\section{Introduction}
There are certain contracts called options which are negotiated in the stock market worldwide. An option is a
contract that gives the buyer the right, but not the obligation, to buy or sell a given asset (a particular stock, an
exchage rate or even an index) at a future time $T$ for a fixed price (or exchange rate or index), called the {\sl strike}
price. If the option is a buy option, its value $C_T$ at expiration time
$T$ is given by:
\begin{equation}
C_T = \mbox{Max} (S_T - K,0)
\label{final}
\end{equation}
where we denote $S_t$ the price of the underlying
asset at time $t$  and $K$ is the strike price. The option price
problem is to find a fair price $C_0 = C(S_0,t=0)$ for the option today.

These contracts are in
principle intended to provide protection to the buyer against moves in the market that can result in large losses.
In a seminal paper, Black and Scholes \cite{BS} showed that, if there is no possibility to make a profit without
taking a certain amount of risk (the non-arbitrage hypothesis), then the price of contracts called options is given
by the solution of a second order differential, the Black-Scholes equation.
One of the basic assumptions of the model is that the returns of the underlying
asset follow a geometric brownian motion:
\begin{equation}
\frac{d S}{S} = \mu dt + \sigma dW
\label{process}
\end{equation}
where $\mu$ is the drift rate of the asset, $\sigma$ is called volatility and $dW$ is a Wiener process with probability
distribution of a normal distribution with zero average and variance $\sigma^2 = dt$, that is $N(0,dt)$.
This stochastic differential equation can be solved resulting in:
\begin{equation}
S_T = S_0 e^{(\mu- \sigma^2/2) T + \sigma \sqrt{T} W}
\end{equation}
with $W \sim N(0,1) $ so that, on average we have:
\begin{equation}
\langle S_T \rangle = \frac{1}{\sqrt 2 \pi}
\int_{-\infty}^{\infty} \; dx \; S_0 e^{(\mu- \sigma^2/2) T + \sigma \sqrt{T} x}
\; e^{-x^2/2} = S_0 e^{\mu T}
\end{equation}

Given the process [\ref{process}], the price of an option is given
by the solution of the Black-Scholes equation:
\begin{equation}
\frac{\partial C}{\partial t} + \frac{1}{2} \sigma^2 S^2 \frac{\partial^2 C}{\partial^2 S}
+ r S \frac{\partial C}{\partial S} - r C = 0
\label{BS}
\end{equation}
with ``initial" condition given by [\ref{final}] and where $r$ is the risk-free interest rate. The trained eye will
recognize [\ref{BS}] as the backward Kolmogorov equation for the
conditional probability corresponding to the process
[\ref{process}] with $\mu$ substituted by $r$.

An alternative way to find the price of an option is to use the
Feynman-Kac theorem to write the solution of the Black-Scholes
equation as an expectation value of the final price at expiration
brought at present value by the risk-free interest rate:
\begin{eqnarray}
C_0 &=& e^{-r T} \int_{-\infty}^{\infty} \; dx \; P(x,t) C_T
\nonumber \\
&=&  e^{-r T} \int_{-\infty}^{\infty} \; dx \; P(x,t)\mbox{Max} (S_T - K,0)
\nonumber \\
&=& e^{-r T} \int_{-\infty}^{\infty} \; dx \; P(x,t)\mbox{Max}
 \left(S_0 e^{(r- \sigma^2/2) T + \sigma \sqrt{T} x} - K,0 \right)
\label{feynman}
\end{eqnarray}
where the probability distribution in this case is symply a normal $N(0,1)$
distribution. Solving the Black-Scholes differential equation [\ref{BS}]or
the expectation value [\ref{feynman}] yields the famous
Black-Scholes pricing formula:
\begin{equation}
C_0 = S_0 N(d_1) - K e^{-r T} N(d_2)
\end{equation}
where $N(x)$ is the commulative probability:
\begin{equation}
N(x) =\int_{-\infty}^{x} \; dz \; \frac{e^{-z^2/2}}{\sqrt{2 \pi}}
\end{equation}
and
\begin{equation}
d_1 = \frac{\ln(S_0/K) + (r + \sigma^2/2) T}{\sigma \sqrt{T}}\;,
\;\;\;
d_2 = d_1 -\sigma \sqrt{T}
\end{equation}

The basic assumption of the Black-Scholes framework is that the
returns of the assets are normally distributed. However, this is
not observed in the market. There is ample evidence that the real
probability distributions of the returns present heavy tails and
asymmetry.  The market usually corrects for this
by introducing a non-constant volatility, the so-called volatility
smile. Figure 1 shows schematically the relation between
the volatility smile and the probability distribution of the
returns. If there is a higher probability of large moves in the
tails, it is compensated by an increase in the volatility
implying a larger price for the option.
This volatility smile, however, bears no relation to the
historical volatility obtained by a time series analysis and is
simply a device to reproduce market data.

\begin{figure}[bth]
\centerline{\epsfxsize=5.in\epsfbox{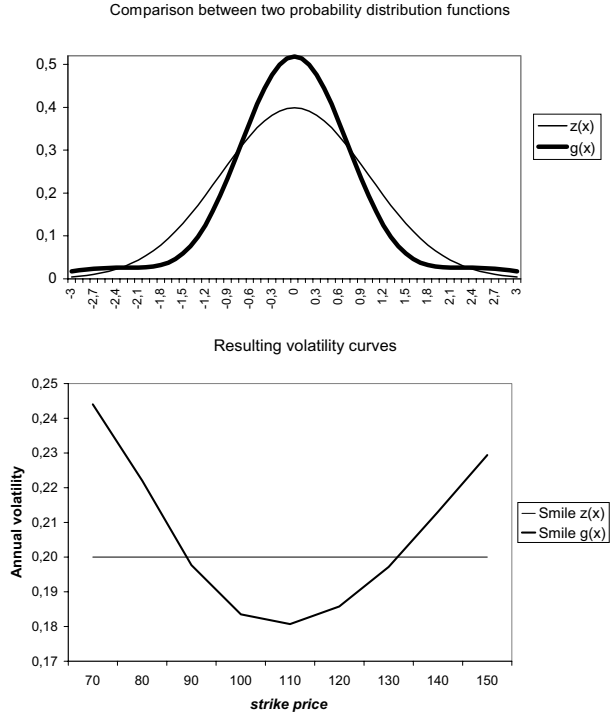}}
\vspace*{-5cm}
\caption{The non-gaussianity of the returns probability distribution is taken
into account by introducing the so-called volatility smile.}
\label{Figure1}
\end{figure}

In this contribution, we study an extension of the
Black-Scholes framework for a non-gaussian distribution.
There is a large number of non-gaussian probability distributions
that could be used. In particular, Lisa Borland recently obtained
a solution for the option price in the context of the Tsallis
distribution \cite{Lisa}. In a more general way, one could
approximate
the real probability distribution by an expansion around the
normal distribution, as suggested by Jarrow and Rudd
\cite{JarrowRudd}. This is the so-called Edgeworth expansion.
In section I we review the Edgeworth expansion, the distribution we
chose to use to model the market returns. In section 2 we derive a
closed form solution for the option price using this distribution
in a risk-free probability measure. In section 3 we test our
results with real market date and show that we can significantly
reduce the volatility smile. We also perform a delta hedge test that shows no
significant improvement over the traditional model. 
Section 5 concludes.

\section{Edgeworth expansion}

In this section we will briefly review the derivation of the
Edgeworth expansion. More details can be found in \cite{Edge}.
Let's assume a series of ${x_1,x_2,...,x_n}$ independent and
identically distributed random variables with mean $\mu$ and
finite variance $\sigma^2$. Defining the random variable
\begin{equation}
X_n = \frac{1}{n} \sum_{i=1}^{n} x_i
\end{equation}
the central limit theorem states that
in the $n \rightarrow \infty$, the random variable $S_n = \sqrt{n}
\frac{(X_n - \mu)}{\sigma} $ approaches a normal distribution $N(0,1)$.
However, we will be interested in the probability distribution before reaching
this formal limit. This can be achieved by a cumulant expansion of
the characteristic function of the distribution $\chi_n(t) = E\left(
e^{i t S_n} \right)$, which for the normal distribution results in
$e^{-t^2/2}$. The characteristic function can be expanded as:
\begin{equation}
\chi_n(t) = \exp\ \left[ -\frac{t^2}{2} + \frac{1}{n^{1/2}} \kappa_3 (i
t)^3 + ...+ \frac{1}{n^{(j-2)/2}} \frac{1}{j!} \kappa_j (it)^j +
... \right]
\end{equation}
where we already used that $E(S_n) = \kappa_1=0$ and
$\mbox{Var}(S_n) = \kappa_2=1$. Performing a Taylor expansion in
$t$ of the characteristic function, collecting terms of the
same order in $n$ and performing the inverse Fourier transform in order to
arrive at the probability distribution, we arrive at the Edgeworth
expansion:
\begin{equation}
g(x) = \left(1 + \frac{\xi}{6} (x^3 - 3x) +
\frac{\kappa-3}{24} (x^4-6x^2+3) + \frac{\xi^2}{72} (x^6 -
15 x^4 + 45 x^2 -15) \right) z(x)
\end{equation}
valid up to order $1/n$  and where we used the skewness $\xi = \kappa_3$ and kurtosis
$\kappa = \kappa_4 +3$, incorporating the factors of $1/n$ in
these parameters. In the equation above $z(x)$ is the gaussian
distribution $N(0,1)$.

The Edgeworth distribution is properly normalized due to
properties of the Hermite polynomials but care must be taken in
the parameters since it may not be positive definite or unimodal.
In Figure 2 we show a graph of the allowed region.

\begin{figure}[bth]
\centerline{\epsfxsize=5.in\epsfbox{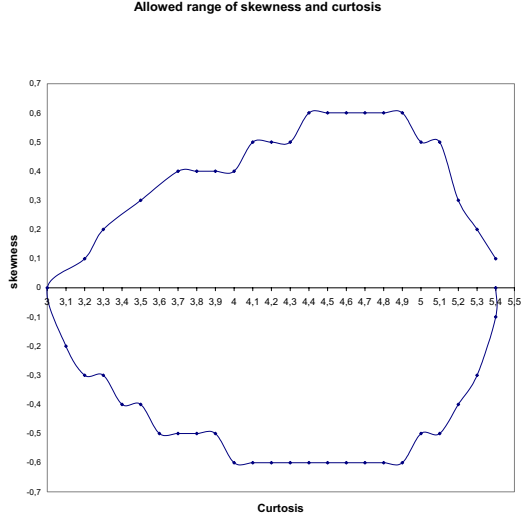}}
\vspace*{-8cm}
\caption{Allowed range of the parameters for a positive definite and unimodal
distribution.}
\label{Figure2}
\end{figure}

\section{Option Pricing with the Edgeworth Expansion}

The distribution of returns of assets in the real markets are known to be non-gaussian, presenting
heavy tails and asymmetry. We will model these returns with the Edgeworth distribution.
The advantage of using the Edgeworth distribution is that it has a
nice analytical form and we will see that it is possible to obtain
a closed, albeit long, form for the fair price of an option on
that asset. This facilitates the use of the results and the tests
we will perform.

The first thing we have to do is to find the risk-free measure in
order to apply equation (\ref{feynman}). In the risk-free measure
the average price of the asset should obey:
\begin{equation}
\langle S_T \rangle = \int_{-\infty}^{\infty} \; dx \; g(x) \;S_0 e^{(\mu - \sigma^2/2) T + \sigma \sqrt{T} x}
= S_0 e^{r T}
\end{equation}
This equation fixes the value of the parameter $\mu$ that must be
used:
\begin{equation}
\mu T = r T - \log\left(1+ \frac{\kappa-3}{24} (\sigma \sqrt{T})^4
+ \frac{\xi}{6}(\sigma \sqrt{T})^3 + \frac{\xi^2}{72}(\sigma
\sqrt{T})^6 \right)
\end{equation}
Notice that one recovers the usual result $\mu = r$ when $\xi=0$
and $\kappa=3$, that is, when we have a gaussian distribution.

The option price is computed by evaluating:
\begin{equation}
C_0^{\mbox {\tiny Edge.}} = e^{-r T} \int_{-\infty}^{\infty} \; dx \; g(x) \mbox{Max}
 \left(S_0 e^{(\mu- \sigma^2/2) T + \sigma \sqrt{T} x} - K,0 \right)
\label{priceEdge}
\end{equation}
This integral can be performed yielding a closed form solution for
the price of the option in the case of the Edgeworth distribution:
\begin{eqnarray}
C_0^{\mbox {\tiny Edge.}} &=& C_0^{\mbox {\tiny BS}} + \\ \nonumber
& & \left( \frac{e^{\mu-rT-x_m^2/2 + \sigma \sqrt{T} x_m}}{72 \sqrt{2 \pi}} S
\left( (\sigma \sqrt{T})^5 \xi^2 + (\sigma \sqrt{T})^4 \xi^2 x_m +
(\sigma \sqrt{T})^3 ( 3 (\kappa-3)+\xi^2 (x_m^2-1)+ \right. \right. \\ \nonumber
& &  \left. \left. (\sigma \sqrt{T})^2 ( 12 \xi - 3 (\kappa-3) x_m +\xi^2 x_m (x_m^2-3) )+
(\sigma \sqrt{T}) ( 12 \xi x_m+3 (\kappa-3)(x_m^2-1)+  \right. \right. \\ \nonumber
& &\left. \left. \xi^2(x_m^4-6x_m^2+3) ) \right) \right)
+ \\ \nonumber
& & \left( \frac{e^{-rT- x_m^2/2}}{72 \sqrt{2 \pi}}
\left( e^{\mu + \sigma \sqrt{T} x_m} S - K \right)
(3 (\kappa-3) x_m (x_m^2-3)+12\xi(x_m^2-1)+ \right. \\ \nonumber 
& & \left. \xi^2 x_m (x_m^4-10 x_m^2+15) ) \right)
+ \\ \nonumber
& & \left( \frac{e^{\mu-rT- \sigma^2 T/2}}{72} S N(d_1)
((\sigma \sqrt{T})^4 3 (\kappa-3) +(\sigma \sqrt{T})^6 \xi^2 + 12 (\sigma \sqrt{T})^3 \xi))
 \right)
\label{closed}
\end{eqnarray}
where $x_m = \frac{\log(K/S_0) - (\mu-\sigma^2/2) T}{\sigma \sqrt{T}}$ is the
minimum value that the random variable can have for a non-zero
integrand in equation (\ref{priceEdge}). Notice that the usual
Black-Scholes result is recovered when $\xi=0$
and $\kappa=3$. In the next section we will use equation
(\ref{closed}) in order to check if the volatility smile can be
reduced.

\section{Edgeworth smile}

In this section we compare the Edgeworth model with the usual
Black-Scholes type of model for the call options of the future
dollar/real exchange rate negotiated at the brazilian BM\&F and
also call options of the S\&P500 index negotiated at the Chicago
Mercantile Exchange. The dollar/real option price are the daily opening
prices in the market.

We want to compare the volatitily smiles obtained from the
Black-Scholes and the Edgeworth model. The volatility smile of the
Black-Scholes model is calculated in the standard way, by
numerically finding the volatility in the model that corresponds
to the market price for a given a strike price.
For the Edgeworth model, we first have to estimate the skewness
and kurtosis parameter. This is achieved by minimizing the total
quadratic error with respect to the model parameters $(\sigma,\xi,\kappa)$:
\begin{equation}
{\mbox Min} \sum \frac{ \left( P_{{\mbox \tiny Market}}
- P_{{\mbox \tiny Edge}} \right)^2}{P_{{\mbox \tiny Edge}}}
\end{equation}
We make sure that the parameters are within the region of validity
of the Edgeworth expansion. Once $\xi$ and $\kappa$ are fixed by
this procedure, we again numerically evaluate the volatility for
different strikes in a given day.

Some representative results are shown in figures 3. We can see
that the smile is significantly reduced when using the Edgeworth
distribution. 

In addition, we have also studied the performance of the Edgeworth model 
compared to the
traditional Black-Scholes models in the delta hedge of portfolio, that is, if
there is a real competitive advantage in using an alternative model in terms of
making money. We concluded that the Edgeworth model does not present a
quantitative gain in the delta hedge test as compared to the Black-Scholes
model.

\begin{figure}[bth]
\centerline{\epsfxsize=5.in\epsfbox{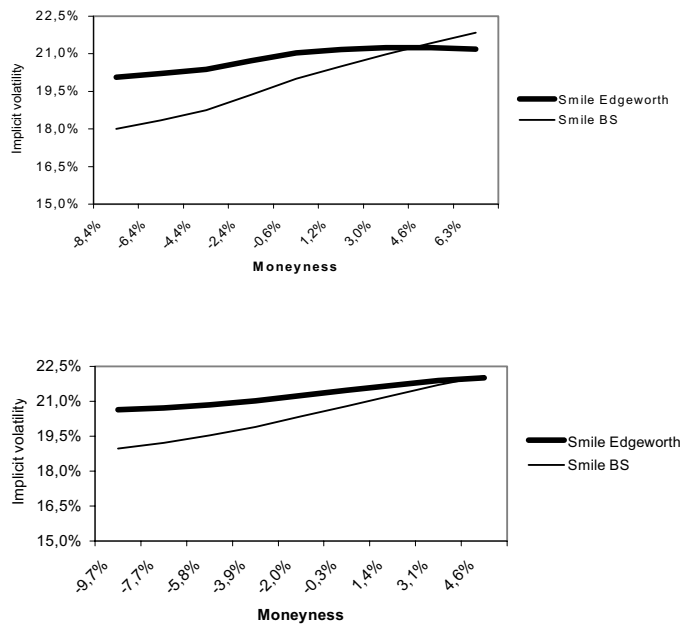}}
\vspace*{-5cm}
\caption{Volatility smile for the Black-Scholes and Edgeworth distributions.
These are smiles obtained from future Real/Dollar exchange rates with two different
expiration dates.}
\label{Figure3}
\end{figure}

\section{Conclusion}

We have studied a non-gaussian extension of the traditional Black-Scholes model,
using the Edgeworth expansion to parametrize the probability distributions of
the asset returns. We derived a closed form expression for the price of an
european option on this asset. We verified with market data that this extension
reduces significantly the volatility smile, implying that the model parameters
are more robust than in the traditional approach. We also performed a delta
hedge test of a given portfolio, where the extension did not show a quantitative
improvement over the Black-Scholes model. This reduces somewhat the interest of
large corporations in a practical use of this extension at this point.

It would be very important to find a microscopic process that would lead to the
Edgeworth distribution. This would provide a solid basis for performing Monte
Carlo simulations for option pricing in this framework.

\section*{Acknowledgements} 
The authors are grateful to Profs. Marcos Eug\^enio da Silva
and Gustavo Athayde for useful comments and suggestions on this work.


\end{document}